\documentclass[3p,times]{elsarticle}

\usepackage{ecrc}
\volume{00}
\firstpage{1}
\journalname{xx}
\runauth{}
\jid{procs}
\jnltitlelogo{xx}
\CopyrightLine{2011}{xx}
\usepackage{amsmath,amssymb,amsfonts}
\newtheorem{myDef}{Definition}

\DeclareMathOperator*{\concat}{\scalebox{1}[1.5]{$\parallel$}}
\usepackage[figuresright]{rotating}
\usepackage{graphicx}
\usepackage{float}  
\usepackage{subfigure}
\usepackage{multirow}
\biboptions{numbers,sort&compress}
\usepackage{color, soul}

\soulregister\cite7
\soulregister\ref7
\sethlcolor{blue}
\setstcolor{red}

\begin{document}

\begin{frontmatter}

\dochead{}

\title{Collaborative Recommendation Model Based on Multi-modal Multi-view Attention Network: Movie and literature cases\tnoteref{label1}}

\author{Zheng Hu$^{a,b}$, Shi-Min Cai$^{a,b,*}$, Jun Wang$^{a,b}$,Tao Zhou$^{a,b}$}
\cortext[cor1]{Correspondence should be addressed to Shi-Min Cai (shimin.cai81@gmail.com)}
\address{a. Comple$\chi$ Lab, School of Computer Science and Engineering, University of Electronic Science and Technology of China, Chengdu 610054, China}
\address{b. Big Data Research Center, University of Electronic Science and Technology of China, Chengdu 610054, China}

\begin{abstract}

The existing collaborative recommendation models that use multi-modal information emphasize the representation of users' preferences but easily ignore the representation of users' dislikes. Nevertheless, modelling users' dislikes facilitates comprehensively characterizing user profiles. Thus, the representation of users' dislikes should be integrated into the user modelling                                                                                                                                                 when we construct a collaborative recommendation model. In this paper,  we propose a novel \textbf{C}ollaborative \textbf{R}ecommendation \textbf{M}odel based on \textbf{M}ulti-modal multi-view \textbf{A}ttention \textbf{N}etwork (CRMMAN), in which the users are represented from both preference and dislike views. Specifically, the users' historical interactions are divided into positive and negative interactions, used to model the user's preference and dislike views, respectively. Furthermore, the semantic and structural information extracted from the scene is employed to enrich the item representation. We validate CRMMAN by designing contrast experiments based on two benchmark MovieLens-1M and Book-Crossing datasets. Movielens-1m has about a million ratings, and Book-Crossing has about 300,000 ratings. Compared with the state-of-the-art knowledge-graph-based and multi-modal recommendation methods, the AUC, NDCG@5 and NDCG@10 are improved by 2.08\%, 2.20\% and 2.26\% on average of two datasets. We also conduct controlled experiments to explore the effects of multi-modal information and multi-view mechanism. The experimental results show that both of them enhance the model's performance.
\end{abstract}


\end{frontmatter}

\section{INTRODUCTION}
In an era of information explosion, recommender systems are the core of many online services. Both users and companies have benefited from the recommender system. For users, it provides a solution to the information overload problem by efficiently directing them to content that aligns with their interests. As for companies, recommender systems serve as a tool to enhance product sales and user engagement, leading to increased revenue. Collaborative filtering is a traditional recommendation strategy widely adopted to establish collaborative recommendation models\cite{su2009survey, EkstrandRK11, lu2012recommender, ShiLH14, KorenRB22, zhang2022survey}. The basic idea behind collaborative filtering is to generate recommendations based on either user or item similarity\cite{SarwarKKR01}. For example, matrix decomposition is a fundamental collaborative filtering algorithm that works by decomposing the user-item interaction matrix into the product of two rectangular matrices with lower dimensions\cite{koren2009matrix}. These matrices represent the user latent vector matrix and the item latent vector matrix. Similarities between users and items are calculated through the dot product of the latent vectors\cite{WangHH22}. Research in recommendation has shifted to developing novel recommender models based on neural networks as a result of deep learning's enormous success in computer vision and language understanding\cite{wang2015collaborative}. So far, deep neural network-based collaborative recommendation models have achieved remarkable results\cite{HeLZNHC17,wang2019neural,wu2022survey}. Knowledge graphs are rich in relationships between entities in the real world, represented as graph structures\cite{JiPCMY22}. In recent years, the increasing research on knowledge graphs has resulted in a proliferation of work utilizing these graph structures to enhance recommendation models\cite{palumbo2017entity2rec,wang2018ripplenet,yang2018knowledge,wang2021learning, WuSZXC23}.

\begin{figure}[htbp]
	\centering
	\includegraphics[width=0.60\textwidth]{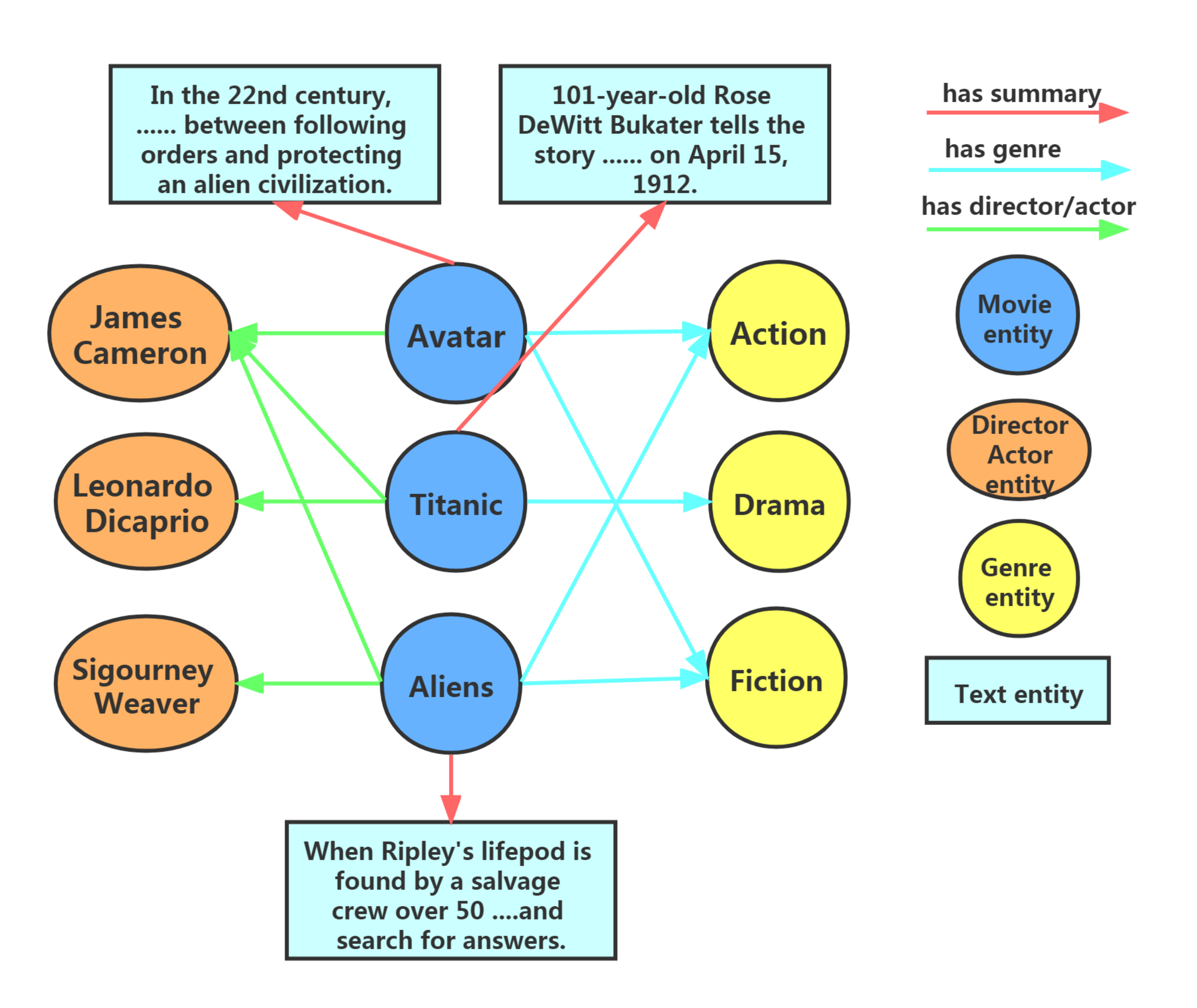}
	\caption{An example of a movie knowledge graph with multiple types of entities. In our work, text entities are used as semantic information. The relationships between entities in the knowledge graph are extracted to serve as structural information.}
	\label{figure1}
\end{figure}

Researchers try to use the rich information in knowledge graphs to improve the performance of recommendation models. For example, Wang et al.\cite{wang2019kgat} construct a collaborative knowledge graph (CKG) by merging the user-item bipartite graph with knowledge graphs. This feat of integration allows the CKG to reinforce the recommendation framework via message propagation across multiple domains. To improve the performance of the recommendation model, Wang et al.\cite{WangHWYL0C21} integrated the fine-grained intent information of user-item interactions discovered in the knowledge graph into the task. However, they usually cannot comprehensively utilize the multi-modal information in knowledge graphs. A simple movie knowledge graph with multiple types of entities (i.e., movie summary, actor, film, and genre) is shown in Fig.~\ref{figure1}. Its multi-modal information not only includes the structural information of connected entities but also involves the semantic information of entities' textual descriptions. For example, for an entity of film, its summary can reflect the movie's content, and its genre can reveal the structural similarity between pairs of films. Such multi-modal information from both the semantic and structural aspects enriches the representation of films, which can improve the accuracy of constructing user and film modelling in a collaborative recommendation model. 
 
The existing recommendation algorithms that use the multi-modal information in various knowledge graphs always consider modelling users' preferences but ignore modelling users' dislikes. For example, Wei et al.\cite{wei2019mmgcn} processed multi-modal information on different user-item interaction graphs and finally gathered the information of different modalities to generate a user representation in the user modelling and an item representation in the item modelling for the recommendation. Sun et al.\cite{sun2020multi} first embedded multi-modal entities in the knowledge graph based on various pre-trained models to obtain their unified representation vectors and performed representation learning to obtain their final representation vectors. These representation vectors of entities are used to merge the user-item interaction graph and the knowledge graph as the collaborative knowledge graph. Then, they aggregated $n$-hop information of the collaborative knowledge graph to generate a user and item representation and performed the dot product to obtain the result. In fact, representing users from multiple views, such as the preference view and dislike view, can more accurately and comprehensively characterize user profiles and improve the recommendation effect.

The previous works employ various types of user behaviour into the recommendation model, such as click, buy, forward, like, etc. \cite{Wu2022FedRec, Ga2010Ex, wu2020neural, Xie2020Deep}. Inspired by this, we try to model users from various types of views. We argue that modelling users' dislikes helps the model build a comprehensive user profile. In addition, structural information and semantic information are often neglected in previous work. We believe that using structural and semantic information can enrich the representation of items and improve the model's performance. In order to address these limitations, we focus on how to effectively utilize the multi-modal information from knowledge graphs and perform the multi-view user representation. With this consideration, we propose a \textbf{C}ollaborative \textbf{R}ecommendation \textbf{M}odel based on \textbf{M}ulti-modal multi-view \textbf{A}ttention \textbf{N}etwork called CRMMAN for short. In general, CRMMAN integrates a novel multi-view mechanism and introduces multi-modal information into the traditional collaborative filtering framework. Specifically, the semantic and structural information of items is extracted from the scene, which indicates the multi-modal information for the item representation. Technically, CRMMAN introduces the multi-view mechanism, which generates the user representations from preference and dislike views based on attention networks. It is worth mentioning that instead of representing users from a single view, we represent each user from both preference and dislike views. The item representations (user representations) are applied for item modelling (user modelling). For an item and a user, we obtain the dot product of the fond user representation and the dislike user representation with the candidate item representation, respectively, of which the weighted sum is denoted as the final prediction result. The contrast experiments are designed based on two benchmark datasets of MovieLens-1M and Book-Crossing, and the evaluation is performed based on two types of metrics.

In all, the contributions of our work are summarized as follows:
\begin{itemize}
	\item We propose a collaborative recommendation model based on the multi-modal multi-view attention network. It comprehensively characterizes the user profile from both preference and dislike views. The multi-view user representation is able to construct user modelling accurately. Furthermore, the semantic and structural information of items extracted from a knowledge graph provides its multi-modal item representation. 
	\item We evaluate the effectiveness of the proposed model based on the designed contrast experiments. The results of contrast experiments suggest that the proposed model outperforms the state-of-the-art method based on knowledge graph and the state-of-the-art method based on multi-modal. 
	\item We present the controlled experimental results of parameter testing in the multi-modal and multi-view attention network. The multi-modal, multi-view, and aggregation layer effects show the mechanism effectiveness of the multi-modal and multi-view attention network. 
\end{itemize}
The remainder of this paper is organized as follows: Section 2 discusses related works. The problem definition is given in Section 3. Section 4 describes the details and training process of our model. Section 5 presents the details of the datasets and the experimental setup. The experimental results and analysis are provided in Section 6. Section 7 is the presentation of conclusions and future work. Tab.~\ref{abbreviation} is the list of abbreviations of this paper.

\begin{table*}[htbp]
	\caption{List of Abbreviations}
	\begin{center}
		\resizebox{0.8\textwidth}{!}{
			\begin{tabular}{|c|c|}
				\hline
				abbreviation&Defination \\ \hline
				KGs & Knowledge Graphs \\ \hline
				MMGCN & Multi-modal Graph Convolution Network \\ \hline
				MGAT & Multimodal Graph Attention Network \\ \hline
				CKAN & Collaborative Knowledge-aware Attentive Network \\ \hline
				KGAT & Knowledge Graph Attention Network \\ \hline
				KGCN & Knowledge Graph Convolutional Networks \\ \hline
				CKE & Collaborative Knowledge Base Embedding Model \\ \hline
				LightGCN & Simplifying and Powering Graph Convolution Network \\ \hline
				CRMMAN & Collaborative Recommendation Model Based on Multi-modal Multi-view Attention Network \\ \hline
			\end{tabular}
		}
		\label{abbreviation}
	\end{center}
\end{table*}

\section{RELATED WORK}
\subsection{knowledge graph-based recommendation}
Knowledge graphs can help to solve the problems of interpretability and cold start of recommender systems. More and more recommendation algorithms based on knowledge graphs have emerged in recent years. Many works regard knowledge graph information as extra information and add it to the original recommendation algorithm. For example, Zhang et al.\cite{zhang2016collaborative} combined collaborative filtering with the text embeddings, knowledge graph structural embeddings, and image embeddings of items to propose an end-to-end recommendation model. Using knowledge-aware convolutional neural networks, Wang et al.\cite{wang2018dkn} integrated the semantic-level and knowledge-level representations of news and generated a knowledge-aware embedding vector. They applied the knowledge-aware embedding vectors as item representations to the collaborative filtering recommendation algorithm.

With the study on message passing mechanism, many works use the message passing mechanism in the knowledge graph for a recommendation. For example, Wang et al.\cite{wang2018ripplenet} drew on Ripple's dissemination, used items as seeds, and conducted Preference Propagation on the item knowledge graph. They believed that the outer items also belong to the user's potential preferences, so when representing the user, they need to be taken into account. Wang et al.\cite{wang2019kgat} pre-processed the knowledge graph using TransR independently. In order to capture the high-level relationships between knowledge graphs and the user-item interaction graph, they build upon the architecture of Attentive Embedding Propagation Layers to recursively propagate embeddings along with high-order connectivity. Finally, they applied these embeddings to collaborative filtering. Cao et al.\cite{cao2019unifying} proposed the Knowledge-enhanced Translation-based User Preference model that jointly learns the representations of users, items, entities, and 
relations. In this way, they captured complementary information from the two tasks to facilitate their mutual enhancements. Wang et al.\cite{wang2019multi} aimed to use the knowledge graph embedding task to assist the recommendation task. They explicitly modelled the high-level interaction between user and item through the cross\&compress unit and automatically controlled the knowledge transfer between the two tasks. Sang et al.\cite{sang2021kg} used a propagating model to learn the embeddings of item entities and user entities in the knowledge graph. Then, the item and user embeddings are fed into an interactive map with hidden convolutional layers to model the complex pairwise correlations between their embedding dimensions explicitly. In this way, the model can discover the high-order interaction information contained in the knowledge graph to improve the performance of the recommendation algorithm. Using auxiliary information from knowledge graphs, Wang et al.\cite{WangHWYL0C21} investigated the intents behind a user-item interaction. They encouraged the independence of various intents for better model capability and interpretability by modelling each intent as an attentive combination of KG relations.

\subsection{Multi-modal Recommendation}
Learning with multiple modalities achieves a more accurate estimate of the latent space representation\cite{HuangDXCZH21}. Researchers have proposed hybrid algorithms that make use of multi-modal information for a recommendation. For example,  Truong et al.\cite{truong2019multimodal} proposed a neural network-based method Multi-modal Review Generation, that simultaneously models rating prediction and comment text. The model used LSTM to process the text in item reviews and used CNN to process pictures in item reviews, and then embedded the two modalities as extra information into collaborative filtering. In order to better capture the user-item information exchange in different modalities, Wei et al.\cite{wei2019mmgcn} designed the Multi-modal Graph Convolution Network(MMGCN) framework. They used the idea of message passing in the graph neural network to spread different modal information in different user-item interaction graphs. Finally, they gathered the information of different modalities to generate a user representation and an item representation for the recommendation. Same as\cite{pezeshkpour2018embedding}, Sun et al.\cite{sun2020multi} regarded different types of information as the relational triples of structured knowledge and proposed a multi-modal graph attention technique called Multi-modal Knowledge Graph Attention Network. They embedded multi-modal entities in the knowledge graph based on various pre-trained models to obtain their unified representation vectors and further performed representation learning to obtain their final representation vectors. These representation vectors of entities were used to merge the user-item interaction graph and the knowledge graph as the collaborative knowledge graph. Then, they aggregated $n$-hop information of the collaborative knowledge graph to generate a user representation and an item representation and performed the dot product to obtain the result. In order to capture various interaction patterns hidden in user behaviours, Tao et al.\cite{tao2020mgat} conducted information propagation within individual graphs of different modalities based on a gated attention graph neural network. The embeddings of users and items from individual graphs of different modalities were fused and applied to the dot-product to make a prediction. Liu et al.\cite{liu2021pre} constructed a homogeneous item graph which provides a unified view of item relations and their side information in multi-modality. Then they proposed the Pre-trained Multimodal Graph Transformer to learn item embeddings. 

\subsection{Multi-modal Representation}
Multi-modal representations can be divided into two categories: joint and coordinated representations. The joint representation projects the information of multiple modalities together into a unified multi-modal vector space. For example, Srivastava et al.\cite{SrivastavaS2012bolzman} proposed a Deep Boltzmann Machine for learning a generative model of multi-modal data. It learns a probabilistic model by sampling from the conditional distributions over each data modality, which extracts meaningful joint representation of multi-modal data. Tan et al.\cite{Tan2019LXMERT} proposed a transformer-based framework, which employed three encoders: a language encoder, an object relationship encoder and a cross-modality encoder to learn the cross-modality representations. Li et al.\cite{Li2020oscar} used object tags detected in images as anchor points. With the anchor point as a reference substance, they applied the self-attention mechanism to text-image pairs to learn the joint representations. The coordinated representations project each modality information to its respective representation space, while certain correlation constraints are satisfied between the projected vectors. For example, Radford et al.\cite{RadfordKHRGASAM2021CLIP} proposed a contrastive framework, which employed an image encoder to obtain the image representations and a text encoder to obtain the text representations. The representations of these two modalities were used to calculate the similarity of image-text pairs. Huo et al.\cite{Huo2021WenLan} designed a two-tower multi-modal pre-trained model, which implicitly models the cross-modal correlation between the image representations and the text representations. Duan et al.\cite{DuanCTYXZC22} encode different modal information into a joint vision-language coding space that is spanned by a dictionary of cluster centers by treating it as different views of the same entity. Through their cluster assignments, they compare positive and negative samples while simultaneously optimizing the cluster centers.

\begin{table*}[htbp]
	\caption{Symbol notion}
	\begin{center}
		\resizebox{0.8\textwidth}{!}{
			\begin{tabular}{|c|c|}
				\hline
				Symbol&Defination \\ \hline
				$L$&length of text sentence \\ \hline
				$s$&semantic information embedding vectors \\ \hline
				$d_e$&BERT word embedding dimension \\ \hline
				$d_h$&semantic information embedding dimension  \\ \hline
				$\mathcal{G}_1$&A KG with muliple types of entities \\ \hline
				$\mathcal{G}_2=\left(\textsl{N},\textsl{E} \right)$&$\mathcal{G}_2$ is an undirected one-part graph,\textsl{N} and \textsl{E} are node set and edge set \\ \hline
				$\alpha$&attention weights in graph attention mechanism \\ \hline
				$P=\{p_1,\dots,p_M\}$&initialization vector set of graph node \\ \hline
				$P'=\{p_1',p_2',\dots,p_M'\}$&intermediate result vector set of graph nodes \\ \hline
				$P''=\{p_1'',p_2'',\dots,p_M''\}$&structural information vector set \\ \hline
				$d_{k''}$&dimension of structural information vector \\ \hline
				$r$&item embedding vector \\ \hline
				$\xi$&user history interactive item set \\ \hline
				$\beta$&weights of self-attention mechanism \\ \hline
				$u_{prefer},u_{dislike}$&user preference representation and user dislike representation \\ \hline
				$click$&item click-through rate \\ \hline
				$c$&embedding vector of the candidate item \\ \hline
				$w1,w2$&learnable parameters, representing the weights of preference and dislike click-through rate \\ \hline
				$W$&learnable parameter matrix \\ \hline
			\end{tabular}
		}
		\label{notion}
	\end{center}
\end{table*}

\section{PROBLEM DEFINITION}
In this section, we give the definitions of our problem and notations. The symbol notations are defined in Tab.~\ref{notion}, which will be used in the following. The basic definitions of our problem include \textit{Knowledge Graph}, \textit{User Implicit Feedback} and \textit{Multi-modal Information}, which will be introduced respectively in the following.

\begin{myDef}
	Recommendation scenarios contain a wealth of knowledge about the items (e.g., item attributes and relationships). We define knowledge graphs (KGs) to represent knowledge about items. A KG $\mathcal{G}_1$ is a directed graph comprised of entity-relation-entity triples (\textit{h}, \textit{r}, \textit{t}), which describes that there is a relation \textit{r} from head \textit{h} to tail \textit{t}. For example, (Avatar, has director, James Cameron) describes the fact that James Cameron is the director of Avatar. It's worth noting that there are many types of entities in the KG, such as text, images, etc.
\end{myDef}

\begin{myDef}
	We assume that in a certain scene, there is a user set $U = \{u_1,u_2,u_3,\dots,u_J\}$ consisting of $J$ users, and an item set $V = \{v_1,v_2,v_3,\dots,v_M\}$ consisting of $M$ items. According to the user's historical interaction records with items, such as clicking news, watching movies, and purchasing items, we can get a user-item interaction matrix $Y\in\mathbb{R}^{J\times M}$. In the $Y$ matrix, $y_{uv}=1$ represents that the user $u$ and item $v$ have interacted and $u$ likes $v$. The remaining items in $Y$ are set to $y_{uv}=0$. It is worth noting that $y_{uv}=0$ has multiple meanings. It not only contains the items $v_{dislike}$ that are recommended to users but not liked by users but also contains items that have no interaction with the user because they have not been recommended. The problem can be defined as follows: Given the user-item interaction matrix $Y$ and the KG of the items, our goal is to predict the user's click-through rate of items and generate a recommendation list for each user based on the click-through rate.
\end{myDef}

	We hope to enrich the representation of items by using multi-modal information to improve the recommendation performance. A KG with multi-modal information is shown in Fig.~\ref{figure1}. We extract two modalities of information from the KG: structural and semantic information. Next, we will give a detailed definition of each modal of information.
\begin{myDef}
	Intuitively, if two entities have a relationship with the same specific entity, then the two entities often have a similar relationship in some way. We try to use the information of the KG to construct an undirected unweighted single-part graph $\mathcal{G}_2=(\textsl{N}$, $\textsl{E})$ to capture this similar relationship as the structural information. $\textsl{N}$ represents the set of nodes and $\textsl{E}$ represents the set of edges. In our method, items are the nodes of the single-part graph, we have $\textsl{N} = V= \{v_1,v_2,v_3,\dots,v_M\}$.		
	For the entities in KG corresponding to items, if there are more than $S_m$ identical related entities between the entity $i$ and the entity $j$, then there is an edge between the corresponding nodes $i$ and $j$, we have $\textsl{E} =\{e_{ij}|\ if\ \mathcal{N}_{i} \cap \mathcal{N}_{j}>S_m\}$.
	$\mathcal{N}_{i}, \mathcal{N}_{j}$ respectively represent the neighbor entity sets of entity $i$ and entity $j$. Structural information reflects the correlation between items, which helps capture users' community preferences for items.
\end{myDef}
\begin{myDef}
	Each item entity has a related text entity in the KG. For example, every movie has a text introduction. We take the text out of the corresponding entity as semantic information. Semantic information can reflect the theme and content of the item, which helps capture the user's preferences. In this paper, texts are segmented into word sequences, denoted as $Semantic_i=\{word_1,word_2,word_3,\dots\}$, representing the semantic information corresponding to item $i$.
\end{myDef}
\begin{figure*}[htbp]
	\centering
	\includegraphics[width=1.0\textwidth]{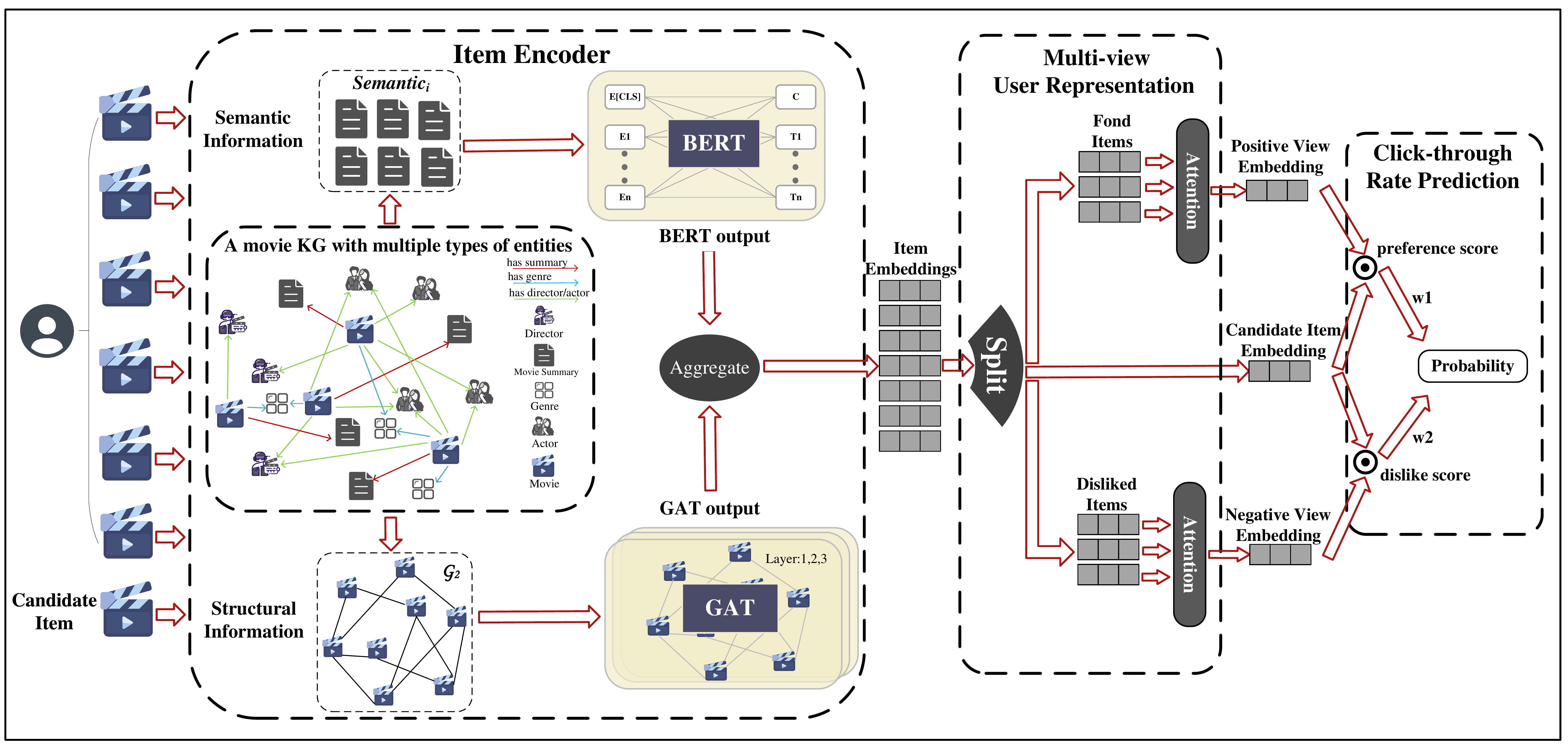}
	\caption{Our model's schematic description, demonstrated with the movie case as an example, is arranged from left to right. As the diagram demonstrates, a diverse range of multi-modal information - including both semantic and structural modalities - is extracted from the knowledge graph and leveraged to encode the items.   The resulting item embeddings are further grouped into items that the user prefers and items the user dislikes, ultimately generating positive and negative views of user representations.}
	\label{framework}
\end{figure*}
\section{METHOD}
In this section, we will introduce in detail the model we proposed. The framework of our model is shown in Fig.~\ref{framework}. As shown in the architecture, the model consists of three parts: 1)Item encoder, which is used to fuse the multi-modal information of the item and embed the item. 2)Multi-view user representation, representing users in preference and dislike views. 3)Click-through rate prediction uses the representation of the candidate item and the user's representations to make a click-through rate prediction. As illustrated in Fig.~\ref{framework}, the input of our model consists of the user's historical interaction items as well as the candidate item, which are then fed into the item encoder. With its access to a knowledge graph encompassing various entity types, the item encoder can encode multi-modal information with semantic and structural modalities into dense latent vectors, thereby encoding the items themselves. The multi-view user representation module takes the item representations as input, divides the items into fond items and disliked items, and then generates the positive view user representation and the negative view user representation using the attention mechanism. The user's positive view embedding, negative view embedding, and the candidate item embedding are input to the click-through prediction module, where the ratings of the two views are computed separately and weighted to obtain the final outcome.

\subsection{Item Encoder}
In reality, items often have more than one modal information. For this reason, we propose an item encoder that can handle multi-modal information. As shown in the left part of Fig.~\ref{framework}, we consider two modalities of information: semantic information and structural information. The item encoder extracts the structural and semantic information corresponding to an item, respectively, and aggregates them as the item embedding.
\subsubsection{Semantic Information Embedding}
We use BERT\cite{devlin2018bert} to encode the semantic information of the text. BERT is a pre-trained Transformer model trained by MLM and NSP methods. For the input text, we use the same tokenizer as BERT to tokenize the text as $ \{t_1,t_2,t_3,\dots,t_L \}$, where $t_1 = [CLS]$, is a special token in the BERT classification task. After that, the tokenized sentence will be passed as input into the BERT model. Specifically, the input layer of BERT constructs the input by summing the token embedding, position embedding and segment embedding. Each kind of embedding is a lookup table with learnable parameters, which can be updated while fine-tuning. The output of BERT is hidden vector matrix $H$ shaped like  $L\times{d_e}$, where $h_i$ represents i-th word embedding and $d_e$ represents the embedding dimension.
\begin{equation}
	\begin{aligned}
	&H=\{t_1',\dots,t_L'\}= BERT(Input) \\
	&Input=TE(tokens)+PE(tokens)+SE(tokens) \\
    &tokens=\{t_1,\dots,t_L\}
    \end{aligned}
\end{equation}
where $TE$ is the token embedding, $PE$ is the position embedding and $SE$ represents the segment embedding. Same as BERT\cite{devlin2018bert}, the input is constructed by summing the three embeddings.

We take out the word embedding $t_1'$ corresponding to $t_1 = [CLS]$. $t_1'$ has condensed the semantic information of the whole sentence. We use $t_1'$ to represent this sentence and input it into a fully connected layer to perform dimensionality transformation. Finally, $s$ with a dimension of $d_h$ is obtained as the extracted semantic information.
\begin{equation}
	s=Wt_1'+b
\end{equation}
where $W \in \mathbb{R}^{d_h\times d_e}$ is the randomly initialized learnable projection matrix, and $b$ is the bias.

In our work, we use Huggingface’s pre-trained BERT-base-uncased model, where the layer of Transformer Encoder $N_t=12$, the dimention is $d_e=768$.

\subsubsection{Structural Information Embedding}
The single-part graph $\mathcal{G}_2$ contains rich item community information. In order to capture this kind of community relationship, an effective method is to embed the nodes of the graph into dense vectors. We use the state-of-art GAT\cite{velivckovic2017graph} algorithm to complete this task. First, we use the SDNE\cite{wang2016structural} algorithm to perform semi-supervised learning on the graph $\mathcal{G}_2$. We take the result as the initialization vector of each node $P= \{p_1,p_2,\dots,p_M\}$, where $p_M\in \mathbb{R}^{d_k}$ and $d_k$ is the embedded dimension. Then, we use a two-layer multi-head attention mechanism to obtain the final node embedding. The attention mechanism is employed to measure the influence of different neighbours on the current node. The node representations learned by the attention mechanism contain the correlation relationships between items, which is helpful for collaborative recommendation. The weight $\alpha$ of our attention mechanism can be expressed as:
\vspace{6pt}
\begin{equation}	
	\alpha_{ij} = \frac{\exp\left(\text{LeakyReLU}\left(\vec{\bf a}^T[{\bf W}p_i\|{\bf W}p_j]\right)\right)}{\sum_{k\in\mathcal{N}_i} \exp\left(\text{LeakyReLU}\left(\vec{\bf a}^T[{\bf W}p_i\|{\bf W}p_k]\right)\right)}
	\vspace{6pt}
\end{equation}
where $\vec{a}\in\mathbb{R}^{2d_{k'}}$, is the weight of a layer of a feed-forward neural network, which is used to realize the attention mechanism. $W\in \mathbb{R}^{d_{k'}\times d_k}$ is a linear transformation matrix used to improve expressive ability. $p_i$ is the target node, $p_j$ is a neighbour node of $i$, and $\mathcal{N}_i$ is the set of neighbour nodes of node $i$. $\alpha_{ij}$ measures the importance of node $j$ to $i$.

We use the multi-head attention mechanism based on the concatenate strategy to obtain the first-level node representation $P'=\{p_1',p_2',\dots,p_M'\}$,$p_M'\in \mathbb{R}^{d_{k'}}$:
\vspace{3pt}
\begin{equation}
	\begin{aligned}
	&p'_i = \concat_{k=1}^K \sigma\left(\sum_{j\in\mathcal{N}_i}\alpha_{ij}^k{\bf W}^kp_j\right) \\
	\end{aligned}
\end{equation}
\begin{equation}
	\sigma(x) =\left\{\begin{aligned}
		&x,\ x\geq0 \\
		&\alpha(e^x-1),\ x<0
		\end{aligned}\right.
\end{equation}
where $\sigma()$ is the ELU\cite{ClevertUH15ELU} activation function, we have $\alpha=1.0$. $\concat$ stands for concatenating operation, $K$ is the head number of the multi-head attention mechanism, $W^k$ is the linear transformation matrix corresponding to the head $k$. It is worth noting that in concatenate mode, the dimension is $d_{k'}=K\times d_k$.

In order to get a better category representation of the node, we perform a multi-head attention mechanism based on the averaging strategy on $P'=\{p_1',p_2',\dots,p_M'\}$ to get the final node embedding $P''=\{p_1'',p_2'',\dots,p_M''\}$,$p_M''\in \mathbb{R}^{d_{k''}}$:
\begin{equation}
	p_i'' =  \sigma\left(\frac{1}{K}\sum_{k=1}^K \sum_{j\in\mathcal{N}_i}\alpha_{ij}^k{\bf W}^kp_j'\right)
\end{equation}
where $	p_i''$ represents the final node embedding. $\sigma()$ is the ELU activation function, $K$ stands for the head number, $\alpha$ is the attention coefficient, $W^k$ is the linear projection matrix. In the averaging strategy, in terms of dimensions, we have $d_{k''}= d_k$.
\subsubsection{Item Embedding}
After getting the semantic information embedding $s_i$ and the structural information embedding $p_i''$ of item $i$, we propose two aggregation methods to obtain the item representation.

\textbf{Concatenate Aggregation:} The concatenate aggregation layer obtains the multi-modal representations of items by concatenating the representations of different modalities. Specifically, we concatenate $s_i$ and $p_i''$ as item representation $r$, $r\in\mathbb{R}^{d_e+d_{k''}}$,
\begin{equation}
	r={s_i}\concat{p_i''}
\end{equation}
where $s_i$ is the semantic modality embedding, and $p_i''$ is the node embedding corresponding to the item $i$. $\concat$ stands for the concatenating operation. The dimension of $r$ is equal to the sum of the dimension of $s_i$ and $p_i''$.

\textbf{Average Aggregation:} When $d_e=d_{k''}$, we can point-wise sum $s_i$ and $p_i''$ and take the average as the item representation $r$,$r\in\mathbb{R}^{d_e}$,
\begin{equation}
	r=\dfrac{1}{2}\left({s_i}+{p_i''}\right)
\end{equation}
where $s_i$ is the semantic modality embedding, and $p_i''$ is the node embedding corresponding to the item $i$. The dimension of $r$ equals the dimension of $s_i$ and $p_i''$.

\subsection{Multi-view User Representation}
In order to model users more comprehensively, each user is represented by two vectors from two views. One vector represents the user's preference, and the other represents the user's dislike. For example, for "Avatar," a sci-fi and romantic movie, when we use two vectors to represent users, users who prefer sci-fi and do not hate romance tend to give a high score, while users who prefer sci-fi but hate romance tend to give a lower score. Nevertheless, when we only use preference user representation, we will predict that these two users both give high scores, which is incorrect.

Intuitively,  the items that interact with the user can reflect the user's preferences and dislikes. Therefore, instead of traditionally set up a separate vector for each user, we use the vector of items that have interacted with the user to represent the user. The historical item set $\xi_u$ of user $u$ can be expressed as:
\begin{equation}
	\begin{aligned}
		&\xi_u^{prefer}=\{v|v\in{V}\ where\ y_{uv}=1\} \\
		&\xi_u^{dislike}=\{v_{dislike} | v_{dislike}\in{V}\ where\ y_{uv}=0\} \\
		&\xi_u=\xi_u^{prefer}\cup \xi_u^{dislike}
	\end{aligned}
\end{equation}
where $\xi_u^{prefer}$ represents the set of items that have interacted with user \textit{u}, and user \textit{u} likes. $\xi_u^{dislike}$ represents the set of items that have interacted with user \textit{u}, but the user does not like.

The attention mechanism has gained popularity in recommender systems in recent years\cite{ZhangYDZ22}. To obtain the user's preference representation and dislike representation, we apply the multi-head self-attention mechanism to the items in $\xi_u^{prefer}$ and $\xi_u^{dislike}$, respectively. Taking preference representation as an example, assuming the size of $\xi_u^{prefer}$ is \textit{z}, the user \textit{u}'s preference representation $u_{prefer}$ can be expressed as:

\begin{equation}
	\begin{aligned}
		&R'=\left(\concat_{x=1}^{X}\left(softmax(\dfrac{RW_{i}^Q\left({RW_{i}^K}\right)^T}{\sqrt{d_{hide}}})RW_{i}^V\right)\right)W^O \\
		&u_{prefer}=Mean\left(R'\right)
	\end{aligned}
\end{equation}
where the Mean() means average the $R'$ matrix along the Y-axis. $R\in\mathbb{R}^{z\times (d_e+d_{k''})}$ represents the matrix of vectors of items that the user \textit{u} likes. $\concat$ stands for the concatenating operation. $X$ represents the number of heads of the multi-head self-attention mechanism. The projections are parameter matrices $W_{i}^Q\in\mathbb{R}^{d_{cat}\times d_{hide}}$, $W_{i}^K\in\mathbb{R}^{d_{cat}\times d_{hide}}$, $W_{i}^V\in\mathbb{R}^{d_{cat}\times d_{hide}}$, $W^O\in\mathbb{R}^{d_{cat}\times d_{cat}}$, where $d_{hide}=\dfrac{d_{cat}}{X}$, $d_{cat} = d_e+d_{k''}$. In the same way, using the item embeddings in $\xi_u^{dislike}$ can get the user $u$'s dislike representation $u_{dislike}$.
\subsection{Click-through Rate Prediction}
This part is used to predict the click-through rate on the candidate item. We apply the dot product method to calculate the click-through rate. Assuming that the item embedding vector of the candidate item \textit{C} is $c$, we use the preference representation and dislike representation of user $u$ to do the dot product with $c$, respectively. Then we apply a weighted summation to get the final click probability:
\begin{equation}
	click=w1\times{c^Tu_{prefer}}+w2\times{c^Tu_{dislike}}
\end{equation}
where $T$ represents the transposition operation, and $w_1,w_2$ represent the weights of preference and dislike, which are two learnable parameters.

\subsection{Model Training}
Inspired by \cite{mikolov2013distributed}, we use the negative sampling strategy to train the model. It is worth noting that negative sampling is the strategy used in training the model, while modelling users' dislikes is a part of model construction. We take the items that the user likes as positive examples. For each positive example, we sample $R$ items from the dislike items that the user has interacted with as negative examples. We take one positive example and $R$ negative examples as a $R+1$ classification problem to train the model. We respectively predict the user's click-through rate $click^+$ for positive example and the click-through rate $click^{-}_1,\dots,click^{-}_R$ for $R$ negative examples, and then calculate the loss:
\begin{equation}
	\begin{aligned}
		&click'^{+}_i=-\log\left({\dfrac{\exp(click^+_i)}{\exp(click^+_i)+\sum_{r=1}^{R}\exp(click^{-}_{i,r})}}\right) 
	\end{aligned}
\end{equation}
\begin{equation}
	\begin{aligned}
		&Loss=\sum_{i\in{U}}click'^{+}_i
	\end{aligned}
\end{equation}
where $click'^{+}_i$ represents the click probability of the i-th positive example after regularization. $click^{-}_{i,r}$ is the click probability corresponding to the r-th negative example negative sampled by the i-th positive example. $U$ is the set of positive examples. Negative sampling can help the model learn the difference between positive and negative examples.

\section{EXPERIMENTAL SETTINGS}
We conduct baseline comparison and ablation experiments on two real-world datasets. In order to enhance the interpretability of our model, we conduct a case study to visualize the semantic information module's results. Hyper-parameter sensitivity experiments are conducted to explore the impact of hyper-parameter settings on the experimental results. In this section, we introduce the experimental settings, and the experimental results, analysis and case study are presented in the next section. Specifically, in this section, we first present the statistics of the datasets and the details of our data preprocessing, then the details of the comparison baselines, and finally, the evaluation scheme and hyper-parameter settings for the experiments.

\begin{figure}[htbp]
	\centering
	\includegraphics[width=1.0\textwidth]{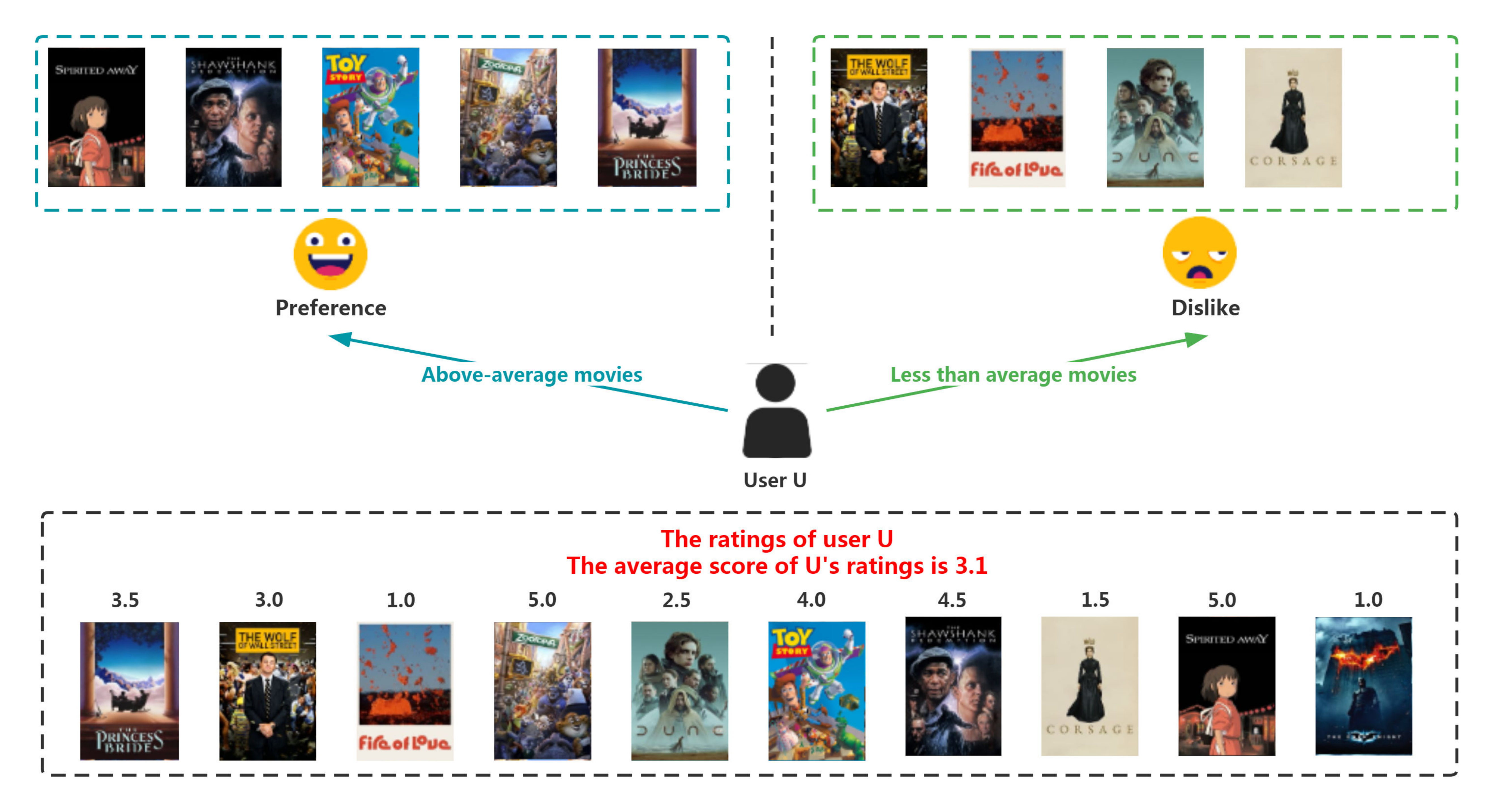}
	\caption{Take the movie scoring scene as an example to explain how to determine user $U$'s preference and dislike. The user $U$ has rated ten movies, and the average score he has given is $3.1$. We think the movies that are above average are the ones he likes, and the movies that are below average are the ones he dislikes. We use the movies he likes to model his interests and the movies he does not like to model his dislikes.}
	\label{Prefer_Dislike}
\end{figure}

\subsection{Dataset}
Our experiment uses the MovieLens-1M\footnote{https://grouplens.org/datasets/movielens/1m/} and Book-Crossing\cite{ziegler2005improving} datasets, which are widely used by recommendation algorithms. The MovieLens-1M dataset contains about one million rating data of 3706 movies from 6040 users. In order to comply with our setting of the question, we converted the user’s rating value to 0 and 1: First, we calculate the average score of each user’s historical rating as $AVG$. Items discretized to 0 which are scored lower than the average ratings given by the user and are therefore considered to be items that the user dislikes, while 1 represents items that the user likes. As illustrated in Fig.\ref{Prefer_Dislike}, for the user $U$’s rating data $rating_{ui}$, we have:
\begin{equation}
	\begin{aligned}	
		&rating^{'}_{ui} = 1 \  if\  rating_{ui}>AVG_u \\
		&rating^{'}_{ui} = 0 \  if\  rating_{ui}<=AVG_u
	\end{aligned}
\end{equation}

\begin{table}[htbp]
	\caption{The Statistic information of the datasets of MovieLens-1M and Book-Crossing.}
	\begin{center}
		\resizebox{0.5\textwidth}{!}{
			\begin{tabular}{|c|c|c|}
				\hline
				\textbf{Datasets}&MovieLens-1M & Book-Crossing \\ \hline
				\textbf{\# users}&5035 & 4715 \\ \hline
				\textbf{\# items}&3659 &  16006 \\ \hline
				\textbf{\# ratings}&969233 & 282913 \\ \hline
				\textbf{\# ratings range}&0.0-5.0 & 0.0-10.0 \\ \hline
				\textbf{\# positive samples}&544323 & 88596  \\ \hline
				\textbf{\# negative samples}&453443 &194317 \\ \hline
				\textbf{\# nodes}&3659 & 16006 \\ \hline
				\textbf{\# edges}&135012 & 1684067 \\ \hline
				\textbf{\# data sparsity}&0.0526 & 0.0033 \\ \hline
			\end{tabular}
		}
		\label{dataset}
	\end{center}
\end{table}

Using this conversion method based on the average score of each user can shield the difference between different users and more accurately reflect users' preferences. For both datasets, we kept items that contained textual descriptions and had at least one edge in $\mathcal{G}_2$. After deleting users with fewer than ten history records, we got a data set of 5035 users, 3659 movies, and about 97k ratings. We used the same rules to process the Book-Crossing dataset and got 282913 scoring records of 16006 books by 4715 users. Detailed statistics of the datasets are shown in Tab.~\ref{dataset}.
\subsection{KG Construction}
In order to construct a KG containing movie semantic information and structural information, we crawled the movies' introduction from MovieLens's website\footnote{https://movielens.org/} as the semantic entity. Also, we crawled the movie director, actor, and movie genres as the entities of the KG. We finally built a KG containing 25613 entities. We set $S_m = 2$ to construct a single-part graph with 3659 nodes and 135012 edges as the structural information of the MovieLens-1M dataset.

In order to construct a KG of books, we crawled the reviews of books from the website\footnote{http://www.bookfinder4u.com} as the semantic information entity. The author and publisher of the books are also used as entities in KG. Finally, we get a KG containing 64024 entities. We use $S_m = 1$ to construct a single-part graph with 16,006 nodes and 16,84067 edges as the structural information of the Book-Crossing dataset.

\subsection{Algorithms of Comparison}
To demonstrate the effectiveness of our proposed model, we compared our model with the following baselines, including the state-of-the-art multi-modal method (MMGCN, MGAT), KG-based methods (CKAN, KGAT, KGCN, CKE) and the graph neural network based methods(LightGCN). We will introduce the hyperparameter settings of baselines in the following subsection.
\begin{itemize}
	\item{MMGCN} \cite{wei2019mmgcn} uses graph neural network to transmit messages on the user-item interaction graphs of different modalities and finally merges the information of each modal to make recommendations. 
    \item{MGAT} \cite{tao2020mgat} is a state-of-the-art multi-modal recommendation algorithm. MGAT leverages gated attention mechanism scores the importance weights of different modalities on user preferences. Compared with MMGCN, MGAT can capture the hidden interaction patterns in user behaviours and make more accurate recommendations.
    \item{CKAN} \cite{wang2020ckan} is a state-of-the-art recommendation algorithm based on the KG. It integrates multiple types of information into the KG and uses different schemes to deal with different relationships. It uses the message passing on the KG combined with the attention mechanism for recommending.
    \item{KGCN} \cite{wang2019knowledge} is an end-to-end framework which extends GCN\cite{kipf2016semi} approaches to the KG. KGCN aggregates and incorporates neighbourhood information biasedly when calculating the representation of entities in the KG, which is able to learn the users' potential interests.
    \item{KGAT} \cite{wang2019kgat} is a KG-based model which explicitly models the high-order connectivities in KG by applying graph attention network on the collaborative knowledge graph of user-item and entity-relation.
    \item{LightGCN} \cite{He2020LightGCN} learns user embeddings and item embeddings by propagating information on user-item interaction graph. LightGCN captures the interaction patterns between users and items through a neighbour aggregation mechanism.
    \item{CKE} \cite{zhang2016collaborative} is one of the classical KG-based recommendation methods. It combines text, structural, and visual knowledge based on collaborative filtering.
\end{itemize}

\subsection{Evaluation Scheme and Hyper-parameter Setting}
In our experiment, we randomly select 70\% of each user's interaction history as the training set and the rest as the test set. Since we have discretized the user's interaction history ratings to 0 and 1, the test data for each user in the test set includes positive and negative samples. For all baselines and our model, we rank each user's interaction samples in the test set to calculate top-K metrics. We randomly select the training set and the test set, perform five independent experiments, and then take the average value as the experimental result. We randomly select part of the data from the training set as the validation set for our method and baselines to help adjust the hyperparameters.
In order to verify the recommendation performance on top-K and the model's classification performance, we apply two widely used metrics for evaluation: NDCG@K and AUC. Here, $K$ values are 5 and 10. For our method and all baselines, we predict the click-through rate of the samples in the test set and generate a recommendation list for each user according to the click-through rate. In addition, we calculate the AUC score for all the samples in the test set.

For our model, after adjusting the hyperparameters, we adopted the concatenate aggregation layer. We set the dimensions of semantic and structural embeddings to 256 and the dimensions of user representation and item representation to 512. The number of heads of the first GAT layer is set to 12, and the second layer is set to 2. Considering the different average length of text in the two datasets, we padded all text to 50 words for MovieLens-1M and 55 words for Book-Crossing. Long texts are truncated at the end, and short texts are padded by 0. The hyper-parameter $B$ controls the number of prefered and disliked history records of each user. In this experiment, we set $B$ to 10. That is, ten preference and dislike history records were over-sampled or under-sampled from the user's interaction history for each user. The number of heads of multi-head self-attention in the user representation is explored in $[2, 4, 8]$ for the both datasets. The negative sampling rate $R$ is set to 4. We apply 40\% dropout to each layer in GAT and 30\% dropout to multi-head self-attention to avoid model overfitting\cite{srivastava2014dropout}. We use Adam\cite{kingma2014adam} as optimizer. The batch size is set to 16.

We implement all baseline models with PyTorch according to the original papers. We have referred to the hyper-parameter values documented in the original papers and have made suitable adjustments to account for our datasets. Our reported results reflect those obtained using the optimal hyper-parameters. Further specifics pertaining to the hyper-parameters used can be found in the following. For common hyper-parameters in all baselines, we explore the learning rate of [1e-2, 1e-3, 1e-4, 1e-5], the batch size of [512, 1024, 2048], the $\ell_2$ regularization of [1e-5, 1e-6, 1e-7]. The number of layers is explored in [2, 3, 4] for all graph neural network-based baseline models. For MMGCN\cite{wei2019mmgcn}, we use the text information processed by BERT and the structural information processed by SDNE as the initial representations of text entities and structural entities. For CKAN\cite{wang2020ckan}, the embedding dimension is fixed to 512. Other hyper-parameters use the default settings in the original paper. For KGAT\cite{wang2019kgat}, as suggested in the original paper, the early stopping strategy is performed. Other parameters are the same as the original paper proposed. For KGCN\cite{wang2019knowledge}, we use the sum aggregator as suggested in the original paper. For CKE\cite{zhang2016collaborative}, we implement it as collaborative filtering plus a structural knowledge module in this paper. For MGAT\cite{tao2020mgat}, to comply with our model setting, we set the embedding dimension to 512.

\begin{table}[htbp]
	\linespread{1.1}
	\tiny
	\caption{The results of contrast experiments based on the datasets of MovieLens-1M and Book-Crossing.}
	\begin{center}
		\resizebox{0.9\textwidth}{!}{
			\begin{tabular}{|c|c|c|c|c|c|c|}
				\hline
				
				\multirow{2}{*} {Methods} &\multicolumn{3}{|c|}{MovieLens-1M}   &\multicolumn{3}{|c|}{Book-Crossing} \\ \cline{2-7}
				
				&{AUC}    &  {nDCG@5}   & {nDCG@10}   & {AUC}  &  {nDCG@5}    & {nDCG@10}    \\ \hline
				MMGCN        & 0.6173         & 0.7541         & 0.7602       & 0.5218           & 0.3882          & 0.4680     \\ 	
				MGAT        & \underline{0.6783}         & \underline{0.7664}          & \underline{0.7640}       & 0.5411           & 0.4121          & 0.4915     \\ 	
				CKAN         & 0.6651          & 0.6859          & 0.6984     &0.5560          & \underline{0.6238}          & \underline{0.6595}    \\ 
				KGAT         &0.5925         & 0.7025      &\ 0.7111          &0.5366         &0.5778      &0.6492         \\
				KGCN         &0.6188         &0.6154      &0.6419          &\underline{0.5707}         &0.5639      &0.6283         \\
				LightGCN         &0.6014         &0.7293      &0.7276          &0.5153        &0.5744      &0.6448         \\
				CKE          & 0.6256          & 0.5545         & 0.6097     & 0.5025   &0.5391         & 0.6088      \\    \hline
				CRMMAN*          & \pmb{0.6981} & \pmb{0.7826} & \pmb{0.7840}  & \pmb{0.5779} & \pmb{0.6382} & \pmb{0.6722} \\ \hline
			\end{tabular}
		}
		\label{movie.result}
		\vspace{-0.1in}
	\end{center}
\end{table}

\section{EXPERIMENTAL RESULTS}
In this section, we first analyze the performance of each model. Then we analyze the effects of multi-modality, multi-view and different aggregation methods on the results one by one. We analyze the visualization results in the case study. Finally, we will explain the results of the hyper-parameter sensitivity experiment. In addition, we try to explore the following three questions through controlled experiments:
\begin{itemize}
	\item{Q1}
	What is the significance of using multi-modal information? How much can we improve by using multi-modal information over only single-modal information?
	\item{Q2}
	What is the reason for using multi-view user representation? What are the advantages compared to the single-view user representation?
	\item{Q3}
	What is the effect of the aggregation layer on the model? Why is there a difference between using different aggregation layers?
\end{itemize}
\subsection{Performance Comparison}
The experimental results of each method are shown in Tab.~\ref{movie.result}. From this table, we have the following observations:

\begin{itemize}
	\item CRMMAN has the best performance on all metrics and all datasets, which fully proves the validity of our model. We attribute these improvements to the use of multi-modality and multi-view mechanisms. Semantic and structural information enriches the representations of items and helps model items comprehensively. Multi-view mechanism simultaneously models users' preferences and dislikes. So CRMMAN can get more complete user profiles than other baselines.
    \item Compare the KG-based methods (CKAN, KGAT, KGCN, CKE). In most cases, the attention mechanisms (CKAN, KGAT) are superior to other methods, indicating that the attention mechanism can aggregate information more effectively than other strategies.
    \item The two models, CKE and MMGCN, performed much worse on the Book-crossing dataset than on the Movielens-1M dataset. Because these two methods rely too much on collaborative filtering based on the user-item interaction graph, the Book-Crossing dataset is much sparser than the Movielens-1m dataset. Therefore, the user and item representation vectors cannot be well trained. 
    \item Compare two multi-modal recommendation algorithms, MGAT and MMGCN, in baselines. MGAT outperforms MMGCN on all metrics for both datasets. This indicates that information propagation on the user-item interaction graph based on the attention mechanism is more suitable for recommendation scenarios than graph convolution. It is worth noting that our CRMMAN model performs better than the state-of-the-art multi-modal algorithm MGAT on all datasets.
\end{itemize}

\begin{figure}[H]
	\centering  
	\subfigbottomskip=2pt 
	\subfigcapskip=-5pt 
	\subfigure[Multi-modal effects]{
		\includegraphics[width=0.45\linewidth]{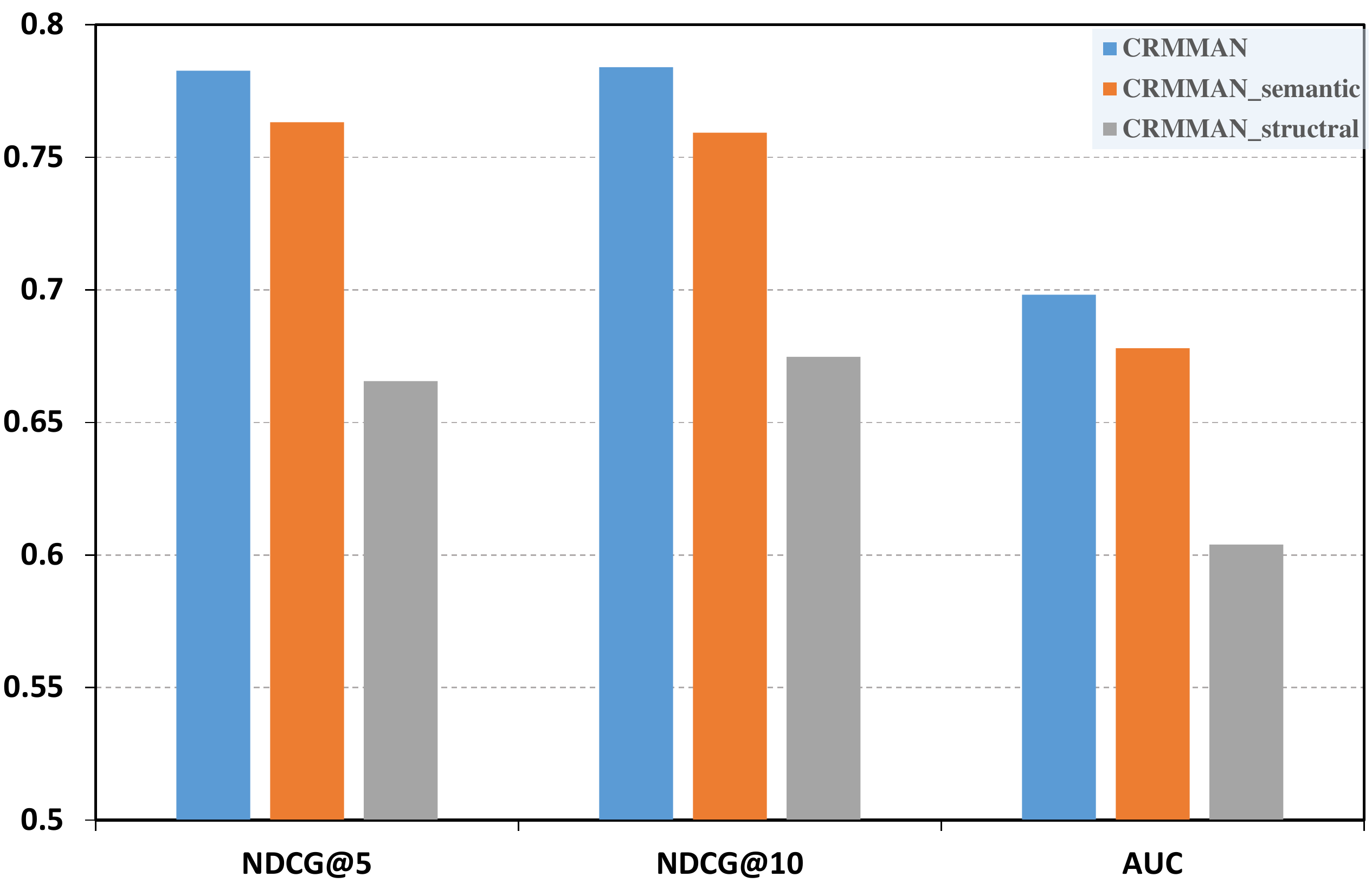}
		\label{multi_modal}}
	\subfigure[Multi-view effects]{
		\includegraphics[width=0.45\linewidth]{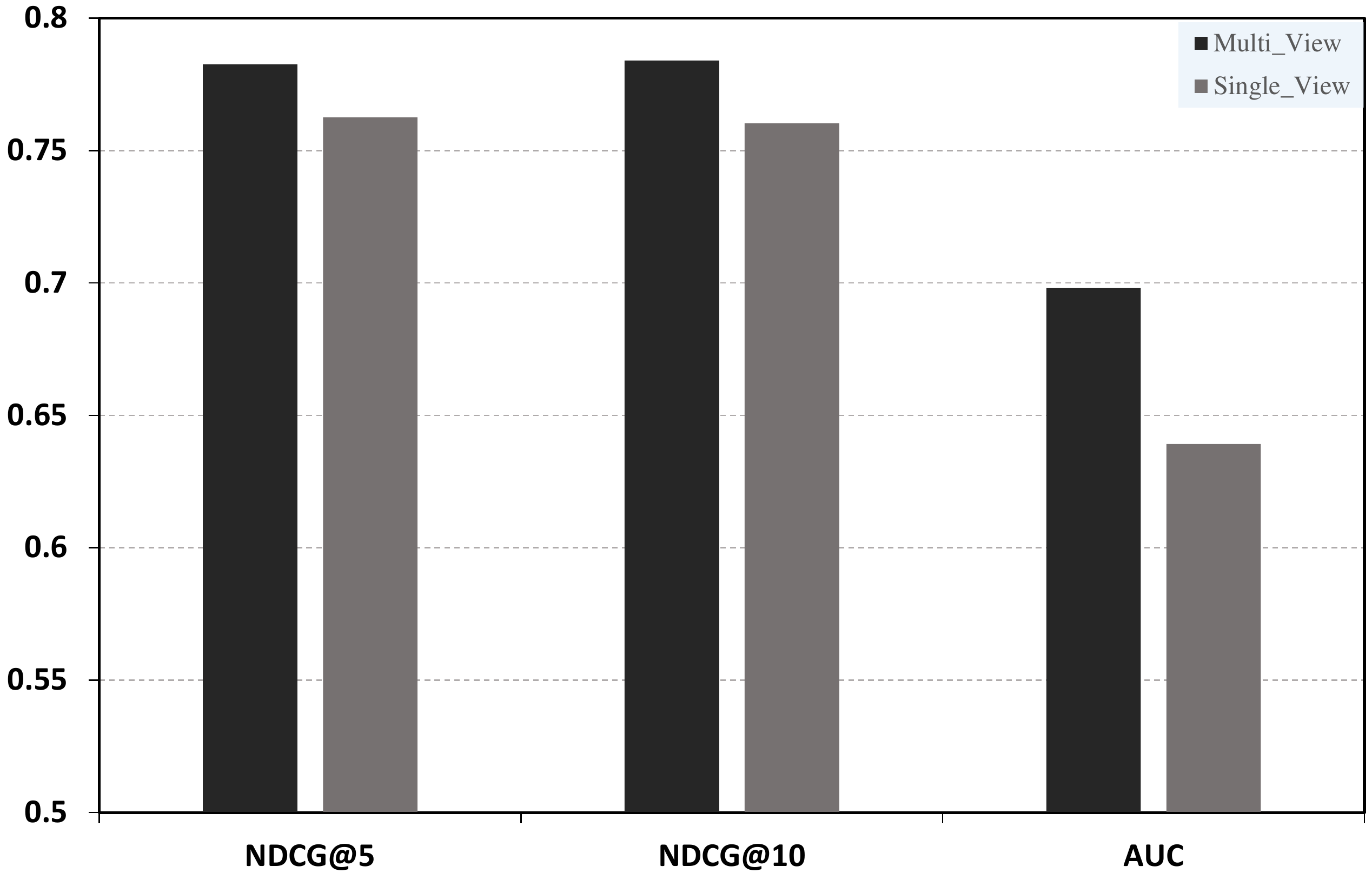}		
		\label{multi-view}}
	\\
	\subfigure[Weights of multi-view]{
		\includegraphics[width=0.45\linewidth]{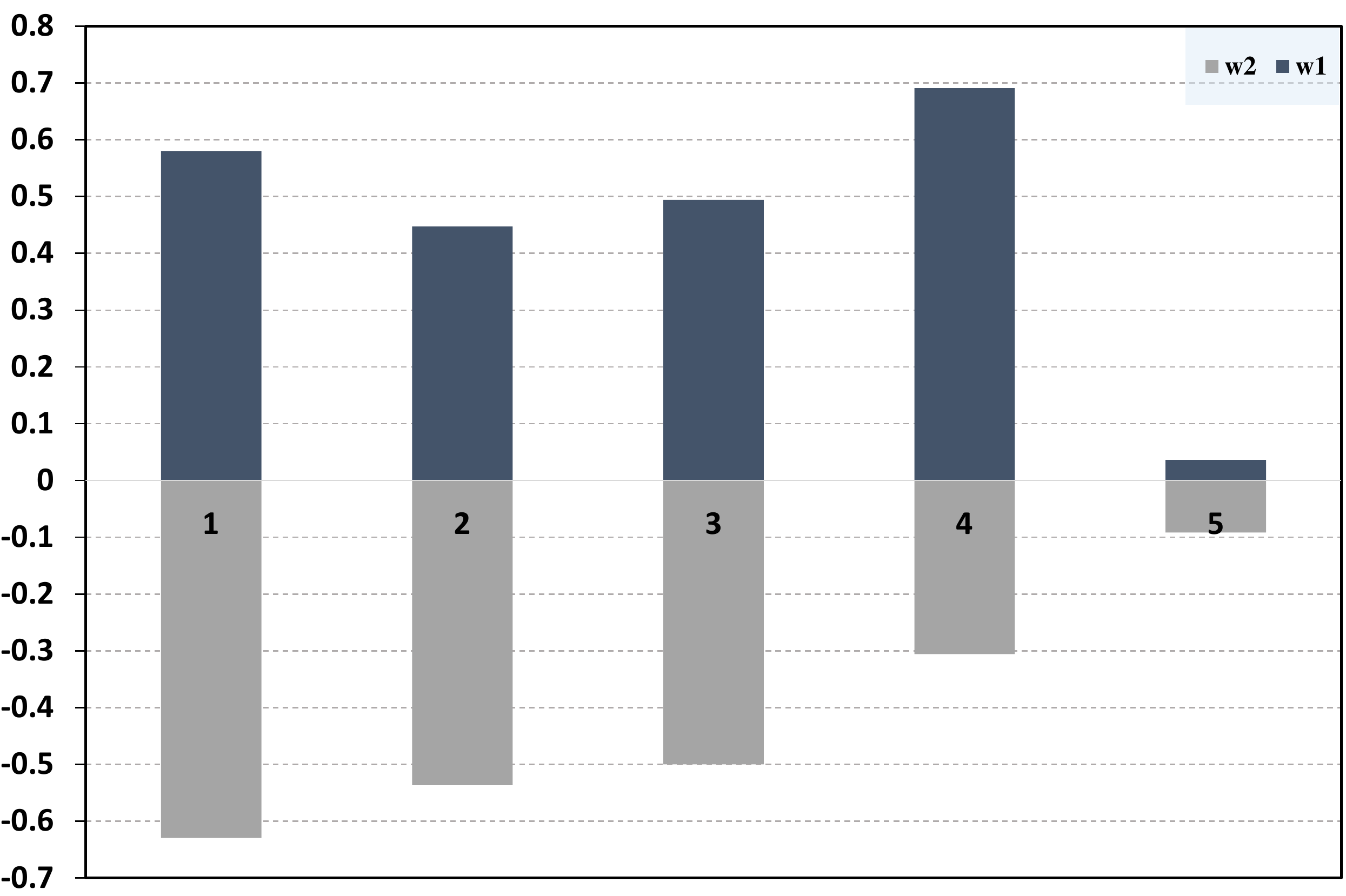}
		\label{figure.view}}
	\subfigure[Aggregation layer effects]{
		\includegraphics[width=0.45\linewidth]{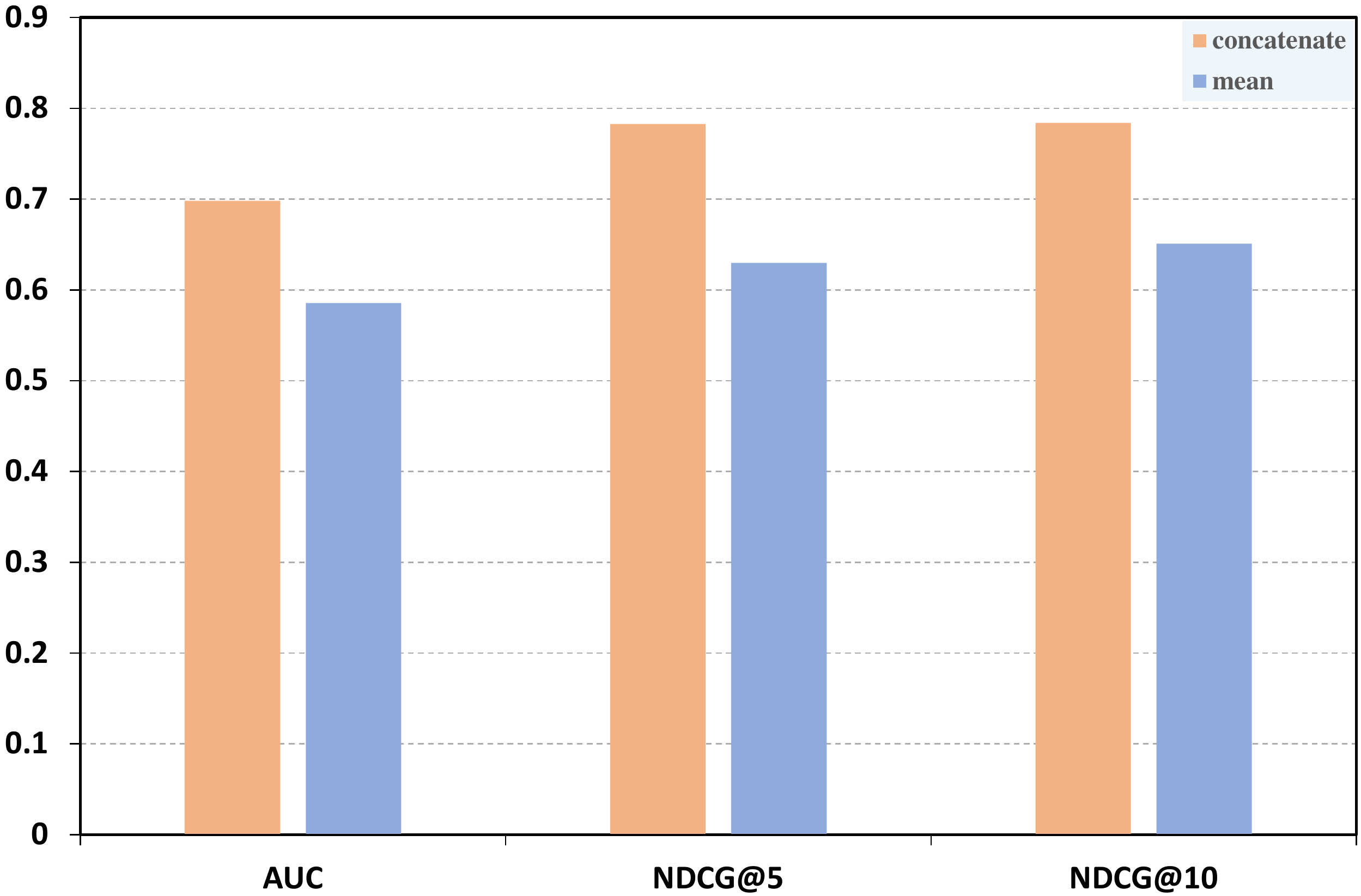}
		\label{figure.agg}}
	\caption{Effects of different mechanisms on results}
\end{figure}

\subsection{Multi-modal Effects\space$\left(\textbf{Q1}\right)$}
In order to explore the influence of different modalities, we conduct experimental comparisons between single and multiple modalities on the MovieLens-1M dataset. Three experiments are carried out independently, using only semantic information, using only structural information, and using both semantic information and structural information. It is worth noting that the model's structures and parameter settings remain the same. The comparison result is shown in Fig.~\ref{multi_modal}. In the figure, CRMMAN is the multi-modal method, CRMMAN\_semantic is the method that only uses semantic modality, and CRMMAN\_structral is the method that only uses structural modality. From Fig.~\ref{multi_modal}, we have the following observations:

\begin{itemize}
    \item As expected, the performance of using multi-modal information is the best in all metrics, which proves the effectiveness of multi-modality for item representation. Compared with the best single-modal results, using multi-modal information improves the three metrics by $\textbf{2.96\%}$, $\textbf{2.54\%}$, $\textbf{3.25\%}$. Intuitively, semantic information helps the model obtain user preferences, and structural information of items can help the model to learn the correlation between items. This observation shows that multi-modal information can comprehensively represent items and thus improve performance. This observation also proves that our model can effectively utilize multi-modal information.
    \item As shown in Fig.~\ref{multi_modal}, the semantic modality plays a more critical role in recommending, which is reasonable because compared to the item community information reflected by the structural modality, the introduction is the information that the user directly touches when choosing a movie. Therefore, semantic information can capture the user's preference for content more accurately.
    \item The structural modal performance is worse than the semantic modal performance. Maybe structural modality can only capture the similarity between items but cannot capture the user's preferences in an all-around way. However, as supplementary information, it can improve the effect of recommendation.
\end{itemize}

\subsection{Multi-view Effects\space$\left(\textbf{Q2}\right)$}
We conduct two independent experiments on the MovieLens-1M dataset to explore the effects of the multi-view mechanism. One experiment uses the multi-view mechanism to model users' preferences and dislikes. The other experiment uses the traditional single-view mechanism and generates only one user vector to model the user. The only variable in the two experiments is whether to use the multi-view mechanism, and the other parameter settings remain the same. The experimental results are shown in Fig.~\ref{multi-view}. In the figure, Multi\_View is the multi-view method, and Single\_View is the method that uses the single-view mechanism. We also visualized the weights of preference and dislike views obtained from five independent repeated experiments, as shown in Fig.~\ref{figure.view}. $w1$ is the weight of the preference view, and $w2$ is the weight of the dislike view. The horizontal axis range 1-5 represents five independent repeated experiments. From Fig.~\ref{multi-view} and Fig.~\ref{figure.view}, we have the following conclusions:
\begin{itemize}
	\item According to Fig.~\ref{multi-view} we find that, as expected, using two views of preference and dislike to model users improves the three metrics by $\textbf{9.22\%}$, $\textbf{2.62\%}$, $\textbf{3.11\%}$ compared to the single view, which is sufficient to prove the effectiveness of the multi-view mechanism. This observation proves that the use of multi-view mechanism can more comprehensively model the user's interests and thus improve the performance of the model.
    \item Comparing the results in Fig.~\ref{multi-view}, it can be found that, compared to the top-K index such as \textbf{NDCG@K}, the multi-view mechanism improves the \textbf{AUC} more significantly. This observation reveals that the multi-view mechanism helps the classification task more than the ranking task.
    \item Through Fig.~\ref{figure.view}, we find that the weights of all preference views are positive, and the weights of all dislike views are negative. As the final output is a weighted sum of the positive and negative views, this observation fully proves that the preference view positively influences the result, and on the contrary, the dislike view negatively influences the result. This observation improves the interpretability of the multi-view mechanism, which also shows its validity of the multi-view mechanism. However, we find that the specific weight values differ for each experiment because the weights are automatically optimized using gradient descent.
\end{itemize}

\subsection{Aggregation Layer Effect\space$\left(\textbf{Q3}\right)$}
In this section, we will study the influence of different aggregation layers. We propose two different aggregation layers, i.e., the concatenate aggregation layer and the average aggregation layer. Keeping parameter settings the same and replacing only the aggregation layer, we test the impact of different aggregation layers on the Movielens-1M dataset. The results are shown in Fig.\ref{figure.agg}.

The results show that the concatenate aggregation layer is better than the average aggregation layer. One possible reason is that the two modal information, semantic and structural information, are in different vector spaces. The average aggregation layer averages the vector element by element, destroying the vector space of different modalities. However, concatenate aggregation can preserve the respective vector spaces of different modalities. Keeping respective vector spaces of different modalities facilitates entirely using the information from different modalities while performing dot-product between the aggregated representations. Therefore, in our model, the concatenation aggregation of the representations of each modality can obtain better results.

\begin{figure}[htbp]
	\centering
	\includegraphics[width=1.0\textwidth, height = 0.25\textheight]{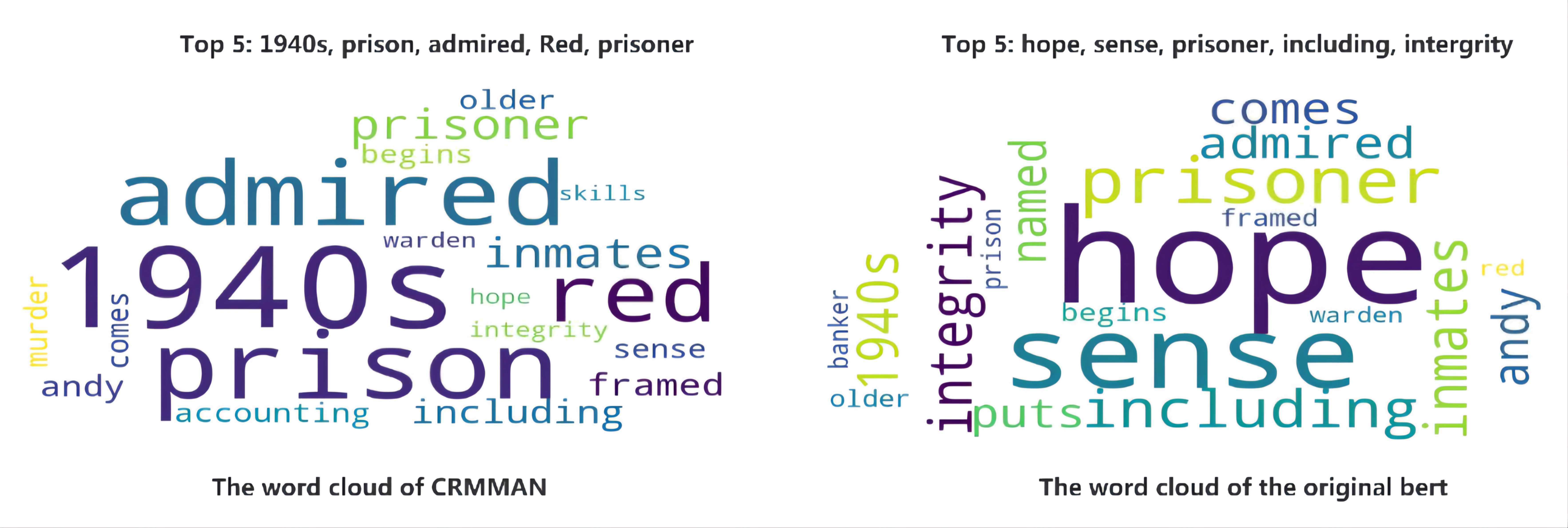}
	\caption{The word cloud on the left is the result of BERT from the trained semantic information processing module in the CRMMAN model. The word cloud on the right is the result of the original BERT without fine-tuning. In the word cloud, the words with great attention from the model are large, and the words with low attention are small. The "Top 5" shows the Top five words that received the most attention.}
	\label{wordcloud}
\end{figure}

\subsection{Semantic Information Embedding Case Study}
To further explore the semantic information embedding part, we take out the trained BERT from the semantic information embedding module in the CRMMAN model, visualize the results, and compare it with the original BERT without fine-tuning. As shown in Fig.~\ref{wordcloud}, we use the introduction of the movie Shawshank Redemption as input to draw word clouds by taking the weights corresponding to the [CLS] token in the attention matrix at the last layer of the BERT models. The introduction of Shawshank Redemption contains 54 words, and we have drawn the word cloud using the top 20 relevant words according to the attention weights. Words with great attention are drawn larger, and words with low attention are drawn smaller.

Compared with the original BERT, we find that the trained BERT from the semantic information module pays more attention to words related to the movie content. For example, the word "1940s", which represents the era of the movie, has been given great attention, and the words "prison" and "prisoner", which are closely related to the content of the movie, have also been given great attention. The original BERT, however, focuses on normal nouns and adjectives like "hope" and "sense". Therefore, the trained semantic information embedding module captures the key content of the movie and can effectively model movies. The visualization results confirm our analysis.

\begin{figure}[H]
	\centering  
	\subfigbottomskip=2pt 
	\subfigcapskip=-5pt 
	\subfigure[Hyper-parameter $B$ influence]{
		\includegraphics[width=0.45\linewidth]{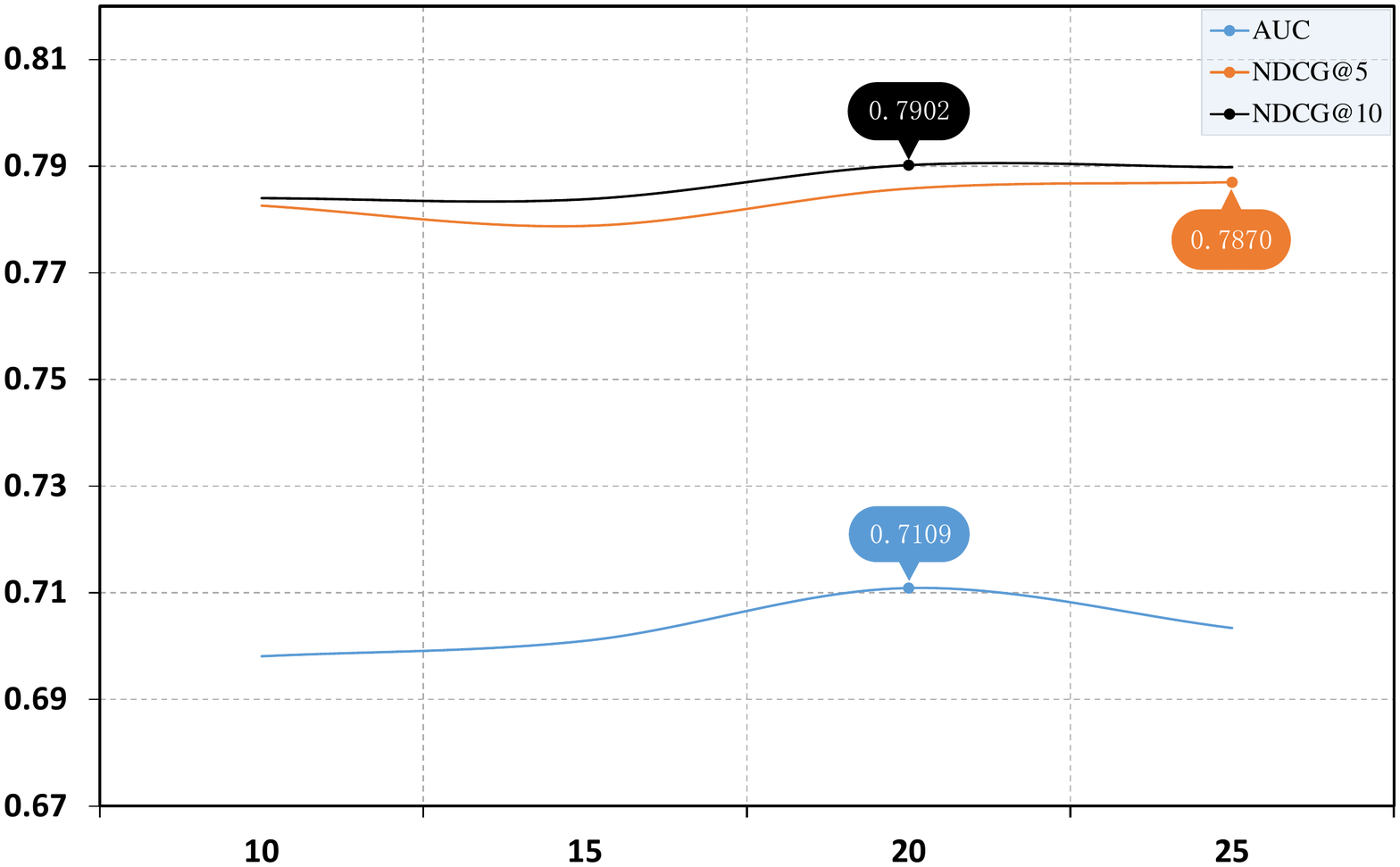}
		\label{C_fig}}
	\subfigure[User embedding dimension influence]{
		\includegraphics[width=0.45\linewidth]{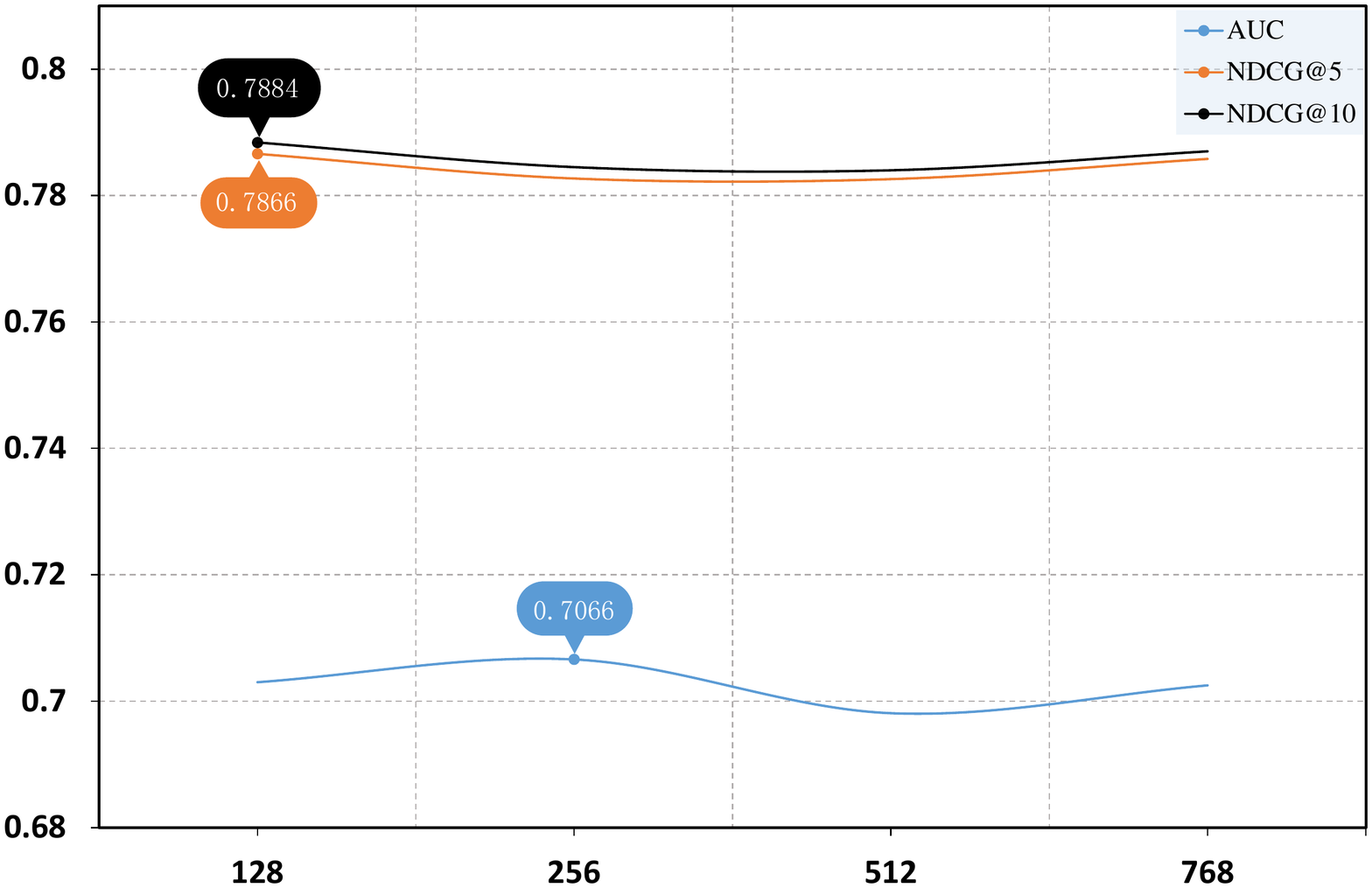}		
		\label{Dim_fig}}
	\caption{These two figures are the results of the hyper-parameter sensitivity experiments. Figure (a) shows the experimental results of hyperparameter $B$, and the X-axis represents different values of $B$. Figure (b) shows the experimental results of the user vector dimension, and the horizontal axis represents the embedding dimension.}
\end{figure}
\subsection{Hyper-parameter Sensitivity Experiment}
In order to explore the effect of hyper-parameter settings on the model, we conducted hyper-parameter sensitivity experiments and analyzed the experimental results. The proposed model has two crucial hyper-parameters: the num of the user history records $B$ and the dimension of the user embeddings. We conducted hyper-parameter sensitivity experiments using the MovieLens-1M dataset. The dataset is preprocessed in the same way as above in the paper. The settings are the same in all experiments except for the specific parameter. The experimental results are shown in Fig.~\ref{C_fig} and Fig.~\ref{Dim_fig}.

The hyper-parameter $B$ determines the number of the user history records used by the proposed model. For each user, the model models the user's interest using $B$ items that the user likes and models the user's dislike using $B$ items that the user hates. We use oversampling and undersampling to ensure that all users have the same number of historical records. We explored the model's performance when $B$ in [10, 15, 20, 25]. The experimental results are shown in Fig.~\ref{C_fig}. The figure shows that the model's performance rises and falls with the increase of the value of hyper-parameter $B$. Apparently, this is a seesaw problem: using more user history records increases the richness of user interests/dislikes. However, a larger value of $B$ would cause more users to oversample from the history records, which causes overfit. It is worth noting that larger values of $B$ also significantly increase model training time. So we tend to use relatively small values of $B$.

The experimental results of the dimension of the user embeddings are shown in Figure.~\ref{Dim_fig}. We explored the model's performance when users embedded dimensions in [128, 256, 512, 768]. Intuitively, larger embedding dimensions have stronger representation power but incur higher time overhead. However, from our experimental results, the user embedding dimension does not have a significant impact on our model. The 128 dimension is the optimal value considering training time and model performance.

\section{CONCLUSION \& FUTURE WORK}
This paper presents a collaborative recommendation model called CRMMAN, which employs multi-modal information to represent items and models users from multiple views. Specifically, we utilize semantic and structural information extracted from KG to represent items. Using multi-modal information avails comprehensively represent items and thus significantly improves recommendation performance. In order to model users in a more granular manner, we design a multi-view user representation mechanism that simultaneously models users' interests and dislikes. In this way, a user's attitude towards an item will be determined by both his preferences and dislikes. We verify CRMMAN with movie and literature recommendation scenarios. Extensive experiments conducted on the MovieLens-1M (movie case) and Book-Crossing (literature case) datasets demonstrate the effectiveness of our model. Our model has achieved an average improvement of $2.08\%$, $2.20\%$ and $2.26\%$ in terms of AUC, NDCG@5 and NDCG@10 compared with the state-of-the-art baselines. Ablation experiments and the case study are conducted to demonstrate the interpretability and effectiveness of the proposed mechanisms.

This work explores the effectiveness of using multi-modal information to make recommendations in movies and books. It can be extended to more areas with multi-modal information, such as short video recommendations and product recommendations in the future. In addition, there are still many exciting works such as \cite{zadeh2017tensor} in exploring the extraction of multi-modal information and the fusion of multi-modal information. In this paper, we explored two aggregation methods to fusion multi-modal information. However, there are still many multi-modal information fusion methods that still need to experiment with. Which multi-modal information fusion method is most suitable for the recommendation scenario needs to be studied in the future. As the diversity of users, we will continue to explore the multi-view representation of users and try to divide users into finer granularity. Only text and graph structural information are used in this paper. We will explore incorporating more modalities into the model in the future. For example, we will try to add gender classification based on computer vision to the model\cite{fekri2019gender}. At the same time, we will explore how to solve the missing modality problem in practice.

\section*{Acknowledgement}
This work is partially supported by the National Natural Science Foundation of China (Grant Nos. T2293771, 61673086, 11975071), and the Ministry of Education of Humanities and Social Science Project (Grant No. 21JZD055). The funders had no role in the study design, data collection, analysis, decision to publish, or preparation of the manuscript.

\bibliographystyle{elsarticle-num}
\bibliography{reference}

\end{document}